Recently, the lubricating effects of oil impregnation in porous solids have become an interesting subject of research. Addition of oil also has a significant effect in modifying the lubrication properties of organic rubbers, such as silicone elastomers (see figure 4 and the related discussions on page 8). In 2008, our group wrote a mini-review on the subject. The theoretical understanding of the subject of soft matter friction has evolved over the years. In that context, many of the ideas presented in this old article may be outdated and somewhat naïve. However, as the effect of oil impregnation in lubrication has received a renewed interest, we post this article with a modified title here hoping that it might be useful to the lubrication community. This is the authors' copy of the manuscript, which was edited and published as a chapter in: "Friction at Soft Polymer Surfaces" Manoj K. Chaudhury, Katherine Vorvolakos and David Malotky, in Polymer Thin Films, Ophelia K. C. Tsui and Thomas P. Russell (eds), World Scientific (2008).

# Friction at Soft Polymer Surfaces: Lubrication Effect of Oil Impregnation

Manoj K. Chaudhury[1], Katherine Vorvolakos[2] and David Malotky[3]

[1]Department of Chemical Engineering
Lehigh University
Bethlehem, PA 18015

[2]FDA Center for Devices and Radiological Health
Office of Science and Engineering Laboratories
Division of Chemistry and Material Science
Silver Spring, MD 20993

[3]The Dow Chemical Co.
Midland, MI 48674

**Abstract** The modes of attachments, detachments and relaxations of molecules of rubbers and gels on solid surfaces are keys to understanding their frictional properties. An early stochastic model of polymer relxations on surfaces was given by Schallamach, which has now evolved in various ways. A review of these developments is presented along with the experimental data that elucidate the kinetic friction of smooth rubber against smooth surfaces. These soft rubbers exhibit various types of instabilities while sliding on surfaces. A few examples of these instabilities are provided.

**Introduction** Friction, in both microscopic and macroscopic systems, is dominated by the processes occurring at or near surfaces, i.e., in a thin film region. The objective of this chapter is to try to understand some of the molecular level processes underlying friction by examining the progress that has taken place over many years with studies involving low modulus polymers. Aside from its tremendous technological importance, crosslinked polymer or rubber is a nice model system for tribological investigations, as its surface and mechanical properties can be easily controlled by the appropriate choice of the molecular weight, crosslinking density, and chemical composition. Another advantage of rubber comes from its low modulus, which allows it to conform well to moderately rough surfaces. A large body of early scientific research in friction was carried out with rubber[1-9]. The questions that intrigued rubber tribologists are: how much of rubber friction is controlled by interfacial and bulk deformation processes, how the dynamics of polymer chains are affected by the interface, what is the mode of relaxation of molecules at the interfaces, to name a few. The pioneering systematic studies of Grosch[10] attempted to answer some of these questions long ago, by making careful measurements of sliding friction of various rubbers against varieties of rough and optically smooth surfaces. His main observation (figure 1) was that the friction coefficient of rubber first increases with velocity, reaches a maximum and then it decreases. In these studies, Grosch[10] discovered a remarkable correlation between friction coefficient and the rheological properties of the rubber. When the velocity corresponding to maximum friction is divided by the frequency corresponding to maximum viscoelastic dissipation, a length scale is found that correlates with the average distance between surface asperities. Based on this observation, Grosch[10] suggested that friction on rough surfaces is dominated by the energy dissipation within a zone in the rubber that is deformed by the surface asperities. As the rough surface slides over a rubber, these zones undergo cyclic compression/deformation cycles, thus dissipating energy proportional to the loss modulus of the rubber. More recently, further extending Grosch's ideas, Persson[23] developed a rather comprehensive theory of rubber friction on rough surfaces. After Grosch[10], Ludema and Tabor[11] carried out careful experiments of rubber friction in both rolling and sliding modes. These authors came to the conclusion that while rolling friction is controlled by the viscoelastic properties of rubber, which gained support from several later studies[12-16], the sliding friction on smooth surfaces does not necessarily correlate well with the viscoelatic properties. They suggested that the friction and slip of polymer chain segments against each other is the main mechanism contributing to sliding friction. On the other hand, Schallamach[17] proposed an adhesion-based theory of rubber friction suggesting that the cyclic processes of extension, detachment and re-attachment of the rubber molecules at the interface are the fundamental mechanisms of friction of rubber on surfaces. Below, we examine these various ideas of rubber friction in some detail.

### Schallamach's Theory of Rubber Friction

Schallamach's idea[10] was that the interfacial bonds responsible for the adsorption of the polymer chains of rubber on a solid surface can always break due to thermal fluctuation. However, when a force is applied, the bond breaking probability increases exponentially with force. By solving a kinetic equation of bond rupture, Schallamach argued that the average force at which a bond breaks increases logarithmically with the rate at which a load is applied to the polymer chain. Interestingly, these basic ideas of Schallamach have also been independently discovered in the biophysics community, where they are frequently used to explain the rate dependent bond rupture forces in various protein-ligand systems[18-20].

Schallamach's[17] theory has two main aspects. According to one aspect of the theory, the force needed to detach a polymer chain from a surface increases with velocity. However, the number of active load-bearing chains decreases with the sliding velocity as well. The force to detach the chain, according to the elementary theory of elasticity, is proportional to the extension of the polymer chain $V\bar{t}$, $\bar{t}$ being the average time a chain remains bonded to a surface before it is detached. This average time is estimated from the following equation:

$$\bar{t} = \int_{o}^{\infty} (\Sigma / \Sigma_o) dt \tag{1}$$

where, $\Sigma_o$ is the total number of polymer chains per unit area. $\Sigma$ is the areal density of working chain at any given time, which is obtained from the following rate equation:

$$\frac{d\Sigma}{dt} = -\frac{\Sigma}{\tau} e^{-U/kT} e^{k_s V t \lambda / kT} \tag{2}$$

where $U$ is the depth of the adsorption potential well, $k_s$ is the spring constant of the rubber molecule, $V$ is the sliding velocity and $\lambda$ is the width of the potential well. The last term in equation 2 suggests that the rate of dissociation of polymer chain from a surface is enhanced due to the reduction of activation energy by the applied force. Integration of equation 2 yields the areal chain density, which can be substituted in equation 1 to obtain $\bar{t}$. The average density of the active load bearing chain is obtained as:

$$\overline{\Sigma} = \Sigma_o \left( \bar{t} / (\bar{t} + \tau) \right) \tag{3}$$

where $\tau$ is the amount of time a chain spends in a detached state before re-attaching to the surface. Combining equations 1, 2 and 3, the frictional force per unit area can be obtained from the sum of all the spring forces: $\overline{\sum} k_s V \bar{t}$. The theory developed by Schallamach is rather successful in providing a qualitative picture of adhesion that captures many important elements of Grosch's observations. Chernyak and Leonov[21] later provided certain improvements to Schallamach's model and proposed the following equation for shear stress:

$$\sigma \approx \sum_o \frac{\int\limits_0^\infty \varphi\left(\frac{r(t)}{\delta}\right) p(V,t)dt}{V(\bar{t}+\tau)}$$

(4)

where $\varphi(r(t)/\delta)$ is the elastic energy stored in the polymer chain, $p(V,t)$ is the transition probability of the polymer chain in going from the bonded to the relaxed state. The numerator of equation 4 is the work done in stretching the polymer chain to the breaking point, while its denominator represents the mean distance traveled by the chain. Using the steady state stochastic model of bonding-debonding processes, Chernyak and Leonov[21] determined the expressions for the transition probability and the time spent in the bonded state in terms of the sliding velocity. Their final expression for the shear stress is:

$$\sigma = \sigma_a (m+1) \frac{u(1+s)\left(1-\exp\left(\frac{-1}{u}\right)-\exp\left(\frac{-1}{u}\right)\right)}{m+1-\exp\left(\frac{-1}{u}\right)}$$

(5)

where $m$ is the ratio of the lifetimes of the polymer chain in the free and bound states at zero sliding velocity, $s$ is the ratio of the viscous retardation time over the lifetime at rest, and $u$ is the dimensionless velocity of sliding defined by $u = \frac{V\tau_o}{\delta a}$, where $\tau_o$ is the lifetime of the polymer chain in the bound state at rest, $\delta$ is the average distance between the polymer body and the wall and $a$ corresponds to the cotangent of the maximum angle the polymer chain subtends on the wall at forced breakup. $\sigma_a = \frac{kT\Sigma_o \delta}{(1+m)R_F^2}$, $R_F$ being the Flory radius of the polymer chain. The resultant shear stress exhibits a bell-shaped behavior with respect to the sliding velocity (Figure 1). The peak stress decreases with the molecular weight, and it exhibits a broad maximum for high molecular weight polymers.

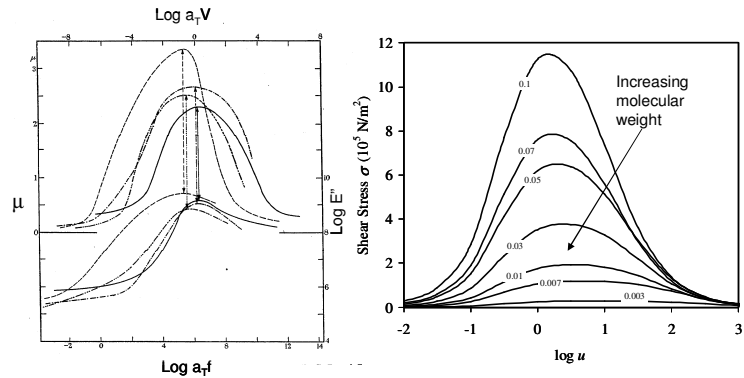

**Figure 1** (left) Friction coefficients of various rubbers against solid surfaces, as obtained by Grosch[10], are plotted against the logarithm of speed shifted by the WLF shift factor. On the same graph, the viscoelastic loss coefficient is plotted against the logarithm of the frequency, also shifted by the WLF factor. (right) Schematic of the shear stress as function of a dimensionless velocity for rubber sliding against a solid surface as predicted by the general theories of Schallamach[17] and Chernyak and Leonov[21].

## Experiments of Rubber Friction with Molecularly Smooth Surfaces

Recently, Vorvolakos and Chaudhury[22,24] studied the friction of a PDMS (polydimethylsiloxane) rubber on a smooth solid surface in the light of the earlier studies by Grosch. The smoothest surface used by Grosch was optical quality glass, which does not guarantee molecular level smoothness. Furthermore, the solid surfaces were rather ill-defined from today's surface science standard; therefore, the effects of surface properties could not be discerned from the older studies of friction. In the new studies, Vorvolakos and Chaudhury[22] used molecularly smooth PDMS (polydimethyl siloxane) elastomers sliding against a molecularly smooth self-assembled monolayer of alkyltricholosilanes or a molecularly smooth polystyrene as a function of the crosslinking density of the elastomer, sliding speed, and temperature.

In these cases, where the interfacial interaction is dominated by van der Waals forces, the frictional force increased with sliding velocity sub-linearly up to a critical velocity, beyond which the rubber exhibited a stick slip behavior (Figure 2). It was also observed that the frictional force decreases with the molecular weight of the rubber, and with temperature, which are qualitatively consistent with the theories of Schallamach, as well as of Chernyak and Leonov[21]. There are certain differences observed with the SAM-coated silicon wafer and polystyrene, which we will discuss later in the chapter.

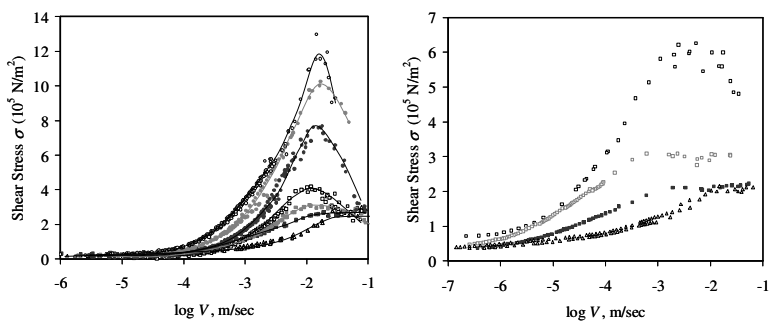

**Figure 2.** (Left) Shear stress of PDMS on SAM-covered Si wafer. The molecular weights of the polymers of decreasing friction correspond to 1.3, 1.8, 2.7, 4.4, 8.9, 18.7, and 52.1 kg/mol, respectively. (Right) Shear stress of PDMS on Polystyrene: PDMS of decreasing friction corresponds to molecular weight 4.4, 8.8, 18.7, and 52.1 kg/mol, respectively.

## Persson and Volikitin's Theory of Rubber Friction on Smooth Surfaces

Recently, Persson and Volikitin[23] questioned whether polymer chains can undergo the cyclic processes of attachment and detachment in a confined geometry as was suggested by Schallamach. They also questioned whether Schallamach's model is suitable for a weakly corrugated potential, where it may be much easier for the chains to slide than to break adhesively. Persson and Volikitin[23] envisioned that the sliding of rubber against a smooth surface is due to the correlated stick-slip motion of the patches of rubber elements which they call stress segments (Figure 3). Applying the method of Langevin dynamics to viscoelastic rubber, Persson and Volikitin[23] argued that the fluctuation stress in rubber segments can be as high as mega-Pascal, which is enough to depin the rubber segment from the surface. When the rubber slides, the stress at the stress segments reach a critical value when a localized slip occurs accompanied with the dissipation of the stored elastic energy in the rubber.

In their analysis, rubber friction turns out to be closely related to the frequency-dependent elastic modulus of the rubber. They recovered (Figure 3) the bell-shaped curve of friction versus sliding velocity as observed in the experiments of smooth surfaces of Grosch[17] as well as of Vorvolakos and Chaudhury[22].

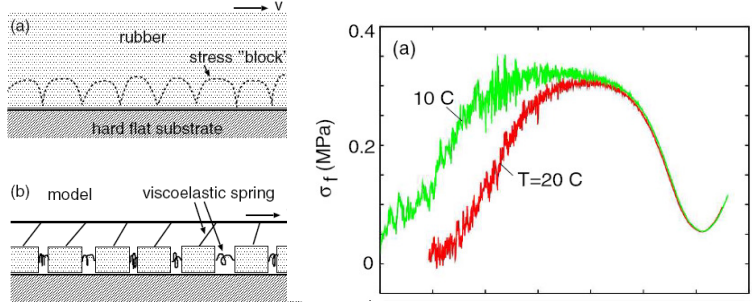

**Figure 3**. fluctuation model[23] of stress segments as proposed by Persson and Volotikin (left). Computed shear stress[23] as a function of sliding velocity is shown on the right.

### Friction Resulting from Stress-Biased Diffusion

Here we describe a model that was used in reference 24, where friction of rubber was thought to be related to the biased diffusion of the polymer molecules in contact with the surface. This model is different from that of Schallamach in the sense that the chains undergo random Brownian motion in contact with the surface, which is biased when a force is applied. For small force, the corresponding Fokker-Planck equation for the dynamics of rubber molecules in contact with the surface is:

$$\frac{Dp}{Dt} = -\frac{D_o}{kT}\frac{\partial(fp)}{\partial x} + D_o\frac{\partial^2 p}{\partial x^2} \tag{6}$$

Here, $p$ is the probability density, $f$ is the force the chain experiences that depends on the chain extension, $D_o$ is the diffusion coefficient. The operator $D/Dt$ indicates a material derivative. At steady state, equation 6 becomes:

$$V\frac{dp}{dx} = -\frac{D_o}{kT}\frac{d(fp)}{dx} + D_o\frac{d^2 p}{dx^2} \tag{7}$$

Solution of equation 7 for a linear elastic polymer chain of spring force $k_s x$ is:

$$p = p_o\, exp\left(\frac{Vx}{D_o} - \frac{k_s x^2}{2kT}\right) \tag{8}$$

The physical significance of equation 8 is that the probability distribution is Gaussian in the absence of any relative sliding. However, when the sliding occurs, the distribution is shifted. The average stress at the interface is:

$$\sigma = \sum\left(\int_{-L}^{L} fp\,dx\right)\Big/\left(\int_{-L}^{+L} p\,dx\right) \tag{9}$$

where $L$ is the maximum extension of the polymer chain. In equation 9, the areal density of load bearing chains $\Sigma$ depends on the statistical segment length $a$ and the number ($N$) of statistical segment per chain as: $\Sigma = 1/\sqrt{N}a^2$. Equation 9 in conjunction with equation 8 reduce to

$$\sigma = \Sigma \zeta_o V \tag{10}$$

where, $\zeta_o = kT/D_o$ is the friction coefficient. This is the low velocity limit of friction, where the frictional stress is areal density of the load bearing chains times the average frictional force a chain experiences due to its diffusive motion. The friction coefficient $\zeta$, in general, has two contributions: one resulting from the surface interactions and the other resulting from viscous friction of the polymer chains with the surrounding chains near the surface. At the simplest level of approximation, the bulk contribution to $\zeta$ is proportional to N, while the surface contribution if proportional to $N^{1/2}$, so the net friction coefficient is:

$$\zeta \sim \zeta_o \left( \sqrt{N} e^{U/kT} + N \right) \tag{11}$$

where $\zeta_o$ is the segmental level friction coefficient in the bulk. If the first term of equation 11 is stronger than the second term, the interfacial shear stress would be $\sigma \approx \dfrac{\zeta_o V}{a^2} e^{U/kT}$, which is independent of the molecular weight (or modulus) of the polymer. At a more simple, but somewhat more general level, the kinetic friction force can be described by an equation[25,26] of the type:

$$f = f_0 \sinh^{-1}(V/V_0) \tag{12}$$

where, $V_o$ is a molecular velocity scale. When $f >> f_o$, the force varies logarithmically with velocity, i.e.

$$f \sim f_0 \ln(V/V_0), \text{ and } \sigma \sim \Sigma f_0 \ln(V/V_0) \tag{13}$$

Thus, while at a low sliding velocity, the shear stress increases linearly with sliding velocity, it varies logarithmically at high sliding velocity.. These results are similar to Schallamach's predictions.

Friction, however, does not keep on increasing indefinitely with velocity. It should saturate when the force reaches the zero temperature limit of the force: $k_x \bar{x} \sim U/\lambda$. However, at the high velocity range, there is a subtle interplay of elastic and frictional forces experienced by the chain. This region may be analyzed using a method proposed by Persson[27] to study the behavior of adsorbed molecules in the high velocity limit using Langevin dynamics. We present here the treatment of Persson[27] that starts with the following Langevin equation:

$$m \frac{d^2 x}{dt^2} + \zeta \frac{dx}{dt} = f - \frac{\partial U(x)}{\partial x} + F \tag{14}$$

Here, $U(x)$ is the depth of the potential well, $f$ is the delta correlated stochastic force and $F$ is the external force. By considering the coordinate of the end group of the chain as $x = Vt + x_a$, where $x_a$ is the fluctuation of the end group position and the shape of the potential well as $U(x) = U(1 - \cos kx)$, equation 14 can be re-written as in equation 15 after expanding the potential with respect to $x_a$.

$$m \frac{d^2 x_a}{dt^2} + \zeta \frac{dx_a}{dt} = f - kU \sin(kVt) - k^2 U u \cos(kVt) + F - \zeta V \tag{15}$$

Persson[27] presented a solution of equation 15 in the high velocity limit as follows:

$$F \approx \zeta V \left( 1 + \frac{U^2}{2m^2 V^4} \right) \tag{16}$$

Thus, at the very high velocity limit, the frictional stress is given by $\sigma = \sum \zeta V$ with $\sum \sim 1/Na^2$ and $\zeta \sim N$; thus, the friction force increases with the molecular weight of the polymer.

Thus, frictional force has three main behaviors. In the low velocity limit, it is dominated by the surface friction coefficient and increases linearly with velocity, but being independent of the molecular weight as a first approximation (the situation may be different if specific interactions, such as H-bonding interactions act at the interface). As the velocity increases, the friction force enters a non-linear domain, where the non-linearity results from the fact that the transition probability from one potential well to the next increases exponentially with force. In this range, the frictional shear stress decreases with the molecular weight of the elastomer. Finally, when the force is much larger than the force of the potential well, friction increases linearly with velocity and molecular weight.

It should, however, be pointed out that in a real rubber, some coupling of interfacial and bulk dynamics may occur. Furthermore, at high sliding velocities, some of the polymer chains may lose contact with the counter surface, or the polymer surface may undergo some roughening, amounting to a decrease of the load-bearing chains with velocity as was envisaged by Schallamach. When this possibility is taken into account, one expects the frictional stress to exhibit an S-shaped curve, where the friction at first increases, then it decreases and finally increases again with sliding velocity. This is also the form predicted by Persson and Volotikin (Figure 3) using their fluctuation model of stress domains and by Leonov[28] after correcting the earlier theory of Chernyak and Leonov by introducing an additional time scale due to Brownian fluctuation:

$$\sigma \cong \sigma_a u \frac{1 - (1 + m + 1/u) \exp(-m - 1/u)}{1 + \gamma m - \exp(-m - 1/u)} \tag{17}$$

where the additional parameter $\gamma$ is the ratio of the free life time of the polymer chain to its characteristic time of Brownian motion on the surface.

### Examination of Experimental Data

The simple analysis presented above suggests that the shear stress is independent of the molecular weight at low sliding velocity. However, with increasing velocity the shear stress should vary as the square root of the elastic modulus. Although there were some hints that friction decreases with the increase of molecular weight in the studies of Newby and Chaudhury[31], the effect was more clearly demonstrated in the studies reported in references 24 and 22. Figure 2 summarizes some of the important results of reference 22. When the sliding speed is in the range of $10^{-7}$ m/s to $10^{-4}$ m/s, shear stress varies very little with velocity. However, when the sliding velocity is in the range of $10^{-4}$ m/s to $10^{-2}$ m/s, shear stress increases logarithmically with the sliding velocity. Then, above V~$10^{-2}$ m/s, sliding becomes unstable followed by stick-slip friction. Using the velocity corresponding to maximum friction ~$10^{-2}$ s and a characteristic length scale ~ 1 nm, one estimates the relaxation time of PDMS chains on the order to be about $10^{-7}$s, which is considerably larger than the viscous relaxation time of the dimethylsiloxane oligomer (~$10^{-11}$s). As a crude approximation, the ratio of these two time scales should be ~ $e^{U/kT}$. This leads to an estimate of $U$ of about 22 kJ/mole. Interestingly, the activation energy of rubber friction from the temperature dependent[22] friction measurements is found to be about 25 kJ/mole, which is close to the above estimate of $U$.

When friction is measured with PDMS on polystyrene surfaces, it is found that the shear stress is somewhat higher than on hydrocarbon SAM. Furthermore, the onset of stick-slip transition for polystyrene occurs at a lower velocity (~$10^{-3}$ m/s). This critical velocity would correspond to a potential well depth of 27 kJ/mole, which corresponds well to the activation energy (27 kJ/mole) of friction obtained from the temperature dependent studies. These differences in friction of PDMS on SAM and polystyrene exemplify the role of interfacial interaction in friction as was pointed out by Ludema and Tabor[11]. Another clear evidence of the role of surface properties on friction was provided by Vorvolakos et al[32] (unpublished results). In these studies, small amounts of uncrosslinked PDMS chains of various molecular weights were mixed with a vinyl end-functional PDMS and the later was crosslinked using the well-known hydrosilation chemistry. The free chains were added in such a small amount (1 weight percent) that the modulus of the network was unaltered. However, the friction of the surface against a polystyrene coated glass sphere decreased dramatically with the molecular weight of the free polymer (Figure 4).

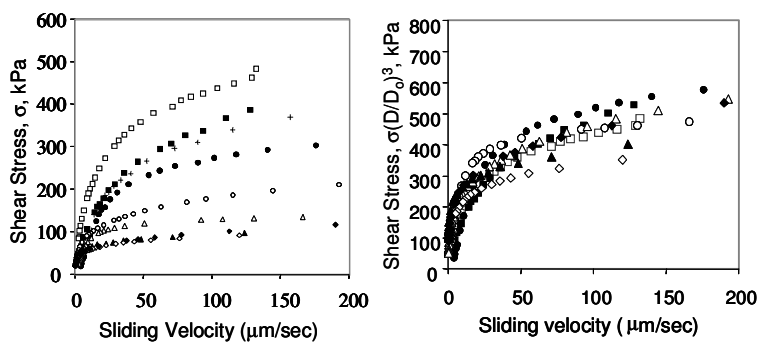

**Figure 4.** (Left) Shear stress of a PDMS elastomer (Mn= 8kg/mole), which was modified by additives of unreactive PDMS of the following molecular weights  0.7 kg/mole (□), 2 kg/mole (■), 14 kg/mole (●), 49 kg/mole (Δ), 116 kg/mole (○), 204 kg/mole (▲), 308 kg/mole (◊), and 423 kg/mole (♦). Clearly, the friction decreases with the increase of molecular weight of additive. (Right) The shear stress data of the left figure are re-normalized by the thickness of the PDMS ($D$) that blooms to the surface. The thickness ($D_o$) of the reference polymer is that of a 0.7 kg/mole additive.

When the molecular weight of the polymer is lower than that of the network itself, a slightly increased friction is observed. It should be mentioned here that the decrease of friction for high molecular weight polymer has also been observed in the melt/solid interfaces[29-30]. Inn and Wang[29] suggested that the stress to desorb a polymer chain from a solid decreases with the molecular weight of the polymer. Hirz et al[30] suggested that the decrease of friction at high molecular weight is related to the reduction of shear stress because of increased thickness. In order to examine the effect of the thickness of the free polymer at the surface, Dave Malotky[32] (unpublished results) developed an AFM based technique. Like the  (remove the space)
previous experiments of Mate et al[33], Malotky found that an AFM tip is pulled towards a substrate by capillary force when it touches the free surface of a thin liquid film. The AFM jump-in distance ($D$) correlates well with the thickness of the film obtained by ellipsometry (Figure 5). However, when the AFM tip is pulled away from the film, it is held by the liquid bridge over a distance (d) that is considerably longer than $D$. These two distances are however linearly correlated, i.e., $D \approx 0.09d + 3.46$. This is a useful relationship as it is much easier to measure $d$ on a soft polymer than $D$, which was used by Malotky to estimate the thickness of liquid contamination films on the surface of a crosslinked polymer.  His results show that the thickness of the layer for small molecular weight PDMS is comparable to their radius of gyration. However, for high molecular weight PDMS, the thickness is considerably smaller than $R_g$. It appears that the large molecular weight PDMS molecules are entangled with the network and they project only a small amount outward in the air. It is noticed that friction decreases with the thickness of this exposed layer. By making a rather crude assumption that the areal density of this exposed portion of the polymer chain varies as $\sim 1/D^2$, we expect that the shear stress of a contaminated network to be proportional to $\sim 1/D^2$. However, when the data of Figure 4 (left) is replotted by multiplying the shear stresses of the polymers with $(D/D_0)^2$ ($D_0$ being the thickness of a reference additive), a good collapse of data is not obtained. In fact the normalization as shown in Figure 4 (right) of the shear stress by multiplying with $(D/D_0)^3$ yields a better result. Although the physics of this scaling is not clear at present, the dramatic decrease of friction with the increase of the molecular weight of the free PDMS additives suggests the role of interface in friction quite convincingly.

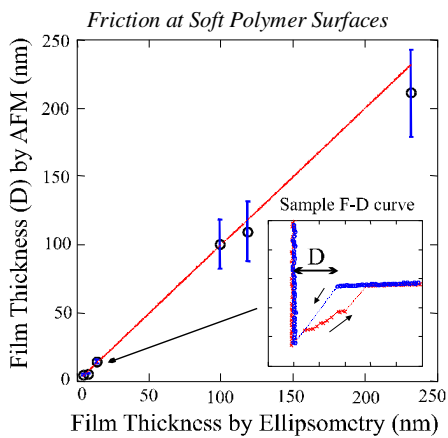

**Figure 5**. Thickness of thin spin-cast PDMS films on silicon wafer as measured by Ellipsometry and AFM jump-in distance. Note that the jump-out distance is much larger than *D*.

Now we return to the subject of investigating the relationship between friction and elastic modulus. First of all, we notice that if the data presented in Figure 2 are normalized by dividing the friction stress with the peak stress, a reasonable master plot is obtained where all the friction data collapse (Figure 6). Furthermore, the peak stress itself increases with the shear modulus of the rubber as $\mu^{0.75}$, where the exponent is found to be somewhat larger than the expected value of 0.5. In Figure 1 as well as in Figure 6 (right) we see that the shear stress of the high molecular weight polymers reaches a plateau value at high velocity without exhibiting the friction maxima observed with the other polymers.

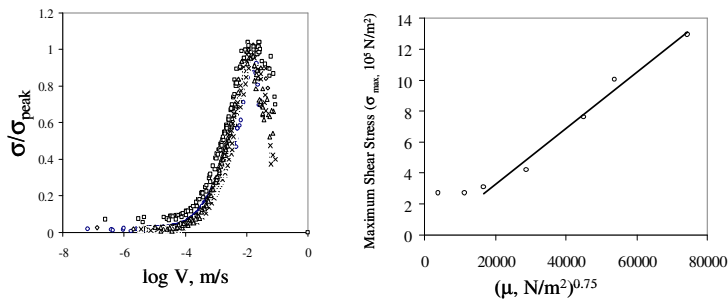

**Figure 6.** (left) Normalization of Shear stress by dividing it with the peak shear stress. Here the data for the PDMS elastomers, which only exhibit peak shear stress and not plateau (see the right figure) are included. (right) Maximum shear stress varies as $\mu^{0.75}$, expect for the high molecular weight polymers[22].

An interesting parallelism has been found in a recent experiment[34] where a rectangular prism of rigid glass was sheared against thin PDMS films of different molecular weights at different velocities (Figure 7).

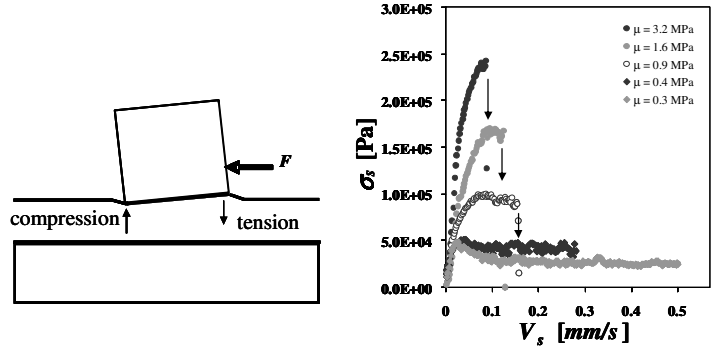

**Figure 7**. Sliding friction of a rectangular prism of hydrophobic glass on 100 μm thick PDMS films of different shear modulus. In this particular configuration, as the glass prism slides, a tensile force is generated at the trailing edge and a compressive force is generated at the advancing edge. When the tensile stress reaches a critical value, the cube comes off the film. The arrows on the right figure indicate when the adhesive fracture occurs. As can be observed on the figure on the right, this kind of failure occurs for high modulus polymers, but not for the polymers of lowest modulus or highest molecular weights. For these highest molecular weight polymers, the normal stress that develops in response to the shear stress is not sufficient to cause adhesive failure. Thus the glass prism keeps on sliding even with increasing speed without coming off the PDMS film. Failure occurs when a defect is encountered. (Adapted from ref 34).

The rigid glass fractures adhesively from the low molecular films at a critical sliding velocity when the normal torque reaches a critical value. For high molecular weight polymers, however, sliding continues even at high velocity without showing any sign of adhesive fracture (except, of course, when a defect is reached). The results of the macroscopic experiments signify that that not enough normal stress is generated for the high molecular weight polymers to detach adhesively. It is plausible that for a related reason at the microscopic level, the areal chain density of the high molecular weight polymers does not decrease to an appreciable value at high velocity to exhibit the typical maximum observed with the low molecular weight polymers. The lack of molecular weight dependence at the high MW polymer may also mean that the entanglement crosslinking density takes over the overall behavior beyond a critical molecular weight.

### Friction in Gel

The subject of gel friction has been elegantly reviewed by Baumberger and Caroli[35]. Here we discuss only a few salient points, which have some resemblance to rubber friction. Gong, Oshada et al[36-42] considered three situations with respect to the friction of a solvated gel against a solid surface. In the first case, the solvent molecules interact more strongly with a solid surface than the polymer chains themselves causing a depletion of the latter at the interface. There is thus a thin solvent layer in between the gel and the solid substrate, the thickness of which depends on the external pressure. By estimating this effective layer thickness that decreases with the applied pressure, Gong et al[41] estimated the viscous friction between the gel and the substrate as follows:

$$\sigma_{rep} / E = \frac{\eta V}{E^{2/3}(k_B T)^{1/3}} \frac{P/E}{1+(P/E)/(1+P/E)^{1/3}} \qquad (18)$$

At low pressure, equation 18 becomes[41],

$$\sigma \cong \frac{\eta VP}{E^{2/3}(k_B T)^{1/3}} \qquad (19)$$

indicating that the shear stress increases with the applied pressure, but decreases with the elastic modulus of the gel.

In the second case, the polymer chains are attracted to the surface. Friction in this regime has two components: one is similar to that of Schallamach, in which the polymer chains undergo cyclic processes of attachment and detachment to the substrate, and the other is hydrodynamic. The shear stress in the regime of adsorption dominated friction is given by the following expression[41].

$$\sigma_{attr} / E = \frac{\eta V}{E^{2/3}(k_B T)^{1/3}} \left[ \frac{1}{2} + \left( 1 + \frac{3}{2} \varepsilon \phi^{-1/2} \right) (1 + P / E)^{1/3} \right] \tag{20}$$

where $\varepsilon$ is a measure of the adsorption energy between a monomer unit and the substrate, and $\phi$ is the volume fraction of the solvent. At high sliding velocity, Gong et al used Schallamach's argument that the number of load-bearing polymer chains spends more time in the detached state than in the attached state causing a decrease of friction with the sliding velocity. However, as the hydrodynamic friction between the gel and the substrate picks up again, friction increases with the sliding velocity. In this limit, the friction stress is given simply in terms of the viscous shear stress. Thus the stress-velocity diagram of gel friction is, in many ways, similar to that of crosslinked rubber. Friction first increases with velocity, it decreases at the intermediate regime and then it increases again (Figure 8). However, unlike the low and the high velocity regimes of rubber friction where the friction is dominated by surface and interchain friction, it is the solvent friction that controls the low and high velocity regimes in the case of gel friction. The case of increasing friction of gel in the high velocity region has been demonstrated recently by Gong et al[42] (Figure 8). However, this region has not yet been clearly identified for rubber.

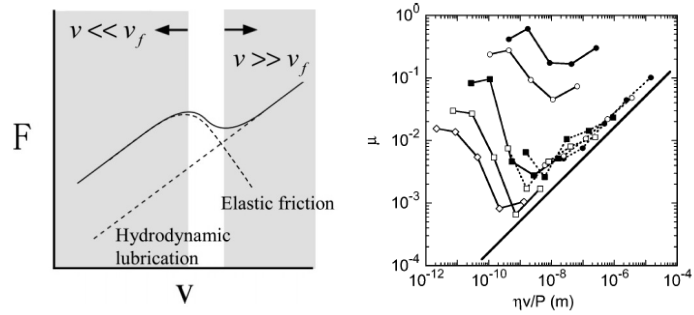

**Figure 8.** Friction of gel showing an S-shaped behavior (left) as expected from the theory[42]. (right) Some high velocity results indeed show the second increase of the friction with velocity[42].

At this juncture, it is worthwhile to mention an interesting elastohydrodynamic effect in gel friction that was suggested by Sekimoto and Liebler[43] as well as Skotheim and Mahadevan[44] in the context of a curved body sliding against a thin soft gel coated substrate (Figure 9).

The main effect comes from the deformation of the gel layer that perturbs the parabolic profile of the gap distances close to the contact region, which affects the hydrodynamic pressure of the liquid that is sheared in the gap. When the perturbation of the profile is taken into consideration in the solution of the standard lubrication approximation of thin film flow in the gap, it is found that the pressure distribution is asymmetric. Integration of this pressure yields the normal force of the following form[43,44]:

$$F \sim \frac{\eta^2 V^2 R^2}{h_o^3} \left( \frac{H_l}{\mu} \right) \tag{21}$$

where $H_l$ is the thickness of the gel, $h_o$ is the gap thickness, and $R$ is the radius of the cylinder. In the case of gel layers sliding against curved

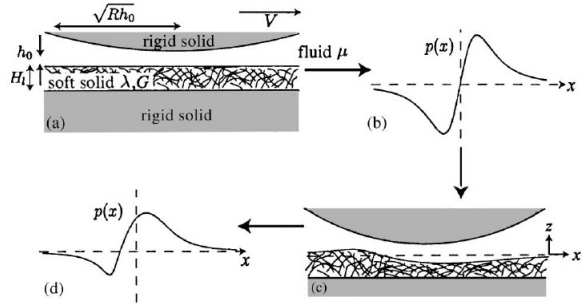

**Figure 9.** This figure shows schematically how asymmetric pressure distribution develops on soft gel when a circular cylinder slides over it (ref. 44)

surfaces, the above normal force can further reduce friction between the surfaces.

### Interfacial Instability and Friction

It has been known for a long time that rubbers exhibit a dynamic instability when sliding on a surface at high velocities. Systematic studies by Grosch showed that friction decreases with sliding velocity when such instabilities are encountered. A nice theoretical analysis of the problem has been provided by Ronsin and Coeuyrehourcq[45]. Recently, Baumberger et al[46,47] made some fascinating observations about the dynamic response at the interface of a sliding gel. When a force is applied at a high rate of deformation of a gel, which is in adhesive contact with a rigid substrate, the force on the gel increases linearly till it reaches a critical state. At this critical state, the force drops to a lower value and homogeneous sliding of gel occurs on the surface. The friction force corresponding to the homogenous sliding exhibits a power law behavior with velocity with an exponent of about 0.4. At low sliding velocity, however, homogeneous sliding does not occur. Instead the force exhibits periodic stick slip dynamics (Figure 10, left). This kind of instability is not in response to the coupling of the system to a weak spring, but it reflects a series of events: breaking of certain numbers of interfacial bonds, transition to a dynamic state, reformation of the bonds and the repetition of the process. Detailed mechanisms of these events were ascribed to the propagation of self-healing slip pulses that nucleate at the trailing edge of the gel slab and propagate to the leading edge. As the slip pulse propagates, the region behind it goes to a stick region, till the pulse reaches the leading edge and another slip pulse is nucleated at the trailing edge to repeat the process (Figure 10).

In confined systems, such as thin films of soft rubber sliding against a rigid substrate, that are prone to instability (Figure 1), bubbles are formed at the trailing edge that propagate towards the leading edge with remarkably high velocities. An approximate analysis describing the propagation of such interfacial bubbles were recently presented in reference 34. The bubble speed can be estimated by balancing the power delivered to the bubbles to the rate of dissipation of energy due to periodic viscoelastic dissipation in the rubber, Chaudhury and Kim[34] found an expression for the bubble velocity as:

$$V_b \sim V_{b0} \left( \frac{\sigma_s}{\mu} \right)^2 \tag{22}$$

where $V_{bo}$ (~ 10 m/s) is a characteristic velocity, and $\sigma_s$ is the applied shear stress. The equation was verified experimentally for the case of a rigid block sliding against thin PDMS elastomers (Figure 11). We end this chapter with the brief discussion of the role of interfacial aging in sliding instability[48] in a somewhat remote analogy to its effect in sliding friction, as discovered by Baumberger et al[46,47]. These experiments were designed to study the energetics of hydrogen bond formation at the interface. To study this effect, an oxidized PDMS rubber hemi-cylinder was brought into contact with a flat silicon wafer coated with a thin layer of grafted PDMS chains. The cylinder was rolled against the flat after different amounts of contact time. From the applied torque, the adhesive fracture energy was estimated, which starts out with a low value similar to that of van der Waals interaction, but it increases fast as contact time progresses and finally reaches to a near plateau value.

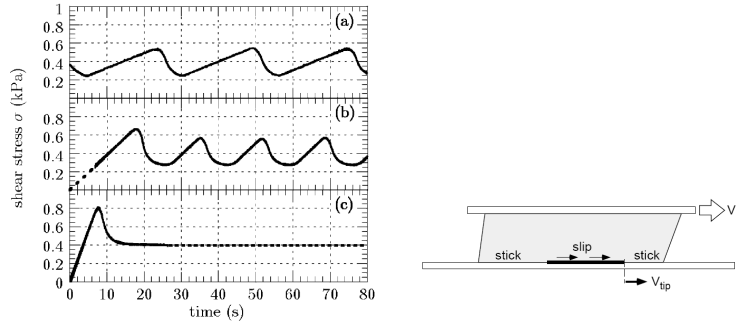

**Figure 10.** (left) Stick-slip friction in soft gel. (right) The propagation of a slip pulse is depicted. As the pulse propagate, the surfaces behind it re-adhere[47].

An interesting observation is made when the hemicylinder is rolled on the surface after some period of contact. For a soft cylinder in contact with a surface, the contact is rectangular with two lines parallel to the long axis of the cylinder. During rolling, one of these lines moves forward before the trailing line starts to move. When the trailing line reaches the boundary of the old and new contact, i.e at the intersection of stronger adhesion due to longer contact time and lower adhesion due to shorter contact time, an elastic instability occurs and the hemi-cylinder rolls forward with considerable speed.

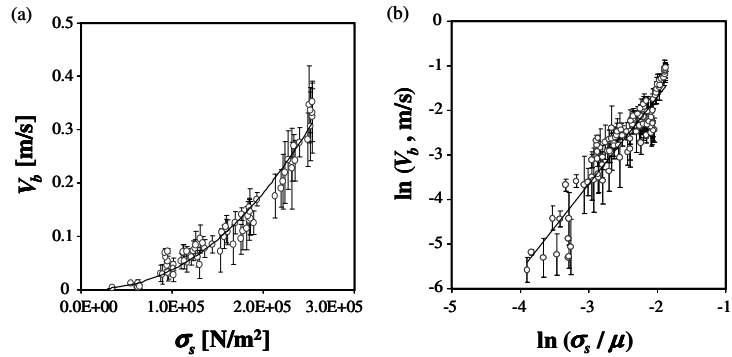

**Figure 11.** Bubble velocity[34] in thin soft elastomers increase with the applied shear stress, but decreases with modulus as predicted by equation 26.

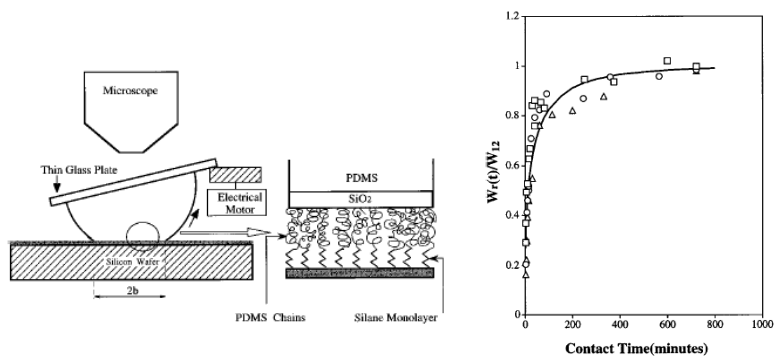

**Figure 12.** (left) Schematic of the measurement of rolling torque of a PDMS hemicylinder against the grafted brush of PDMS[48]. (right) The fracture energy (which is expressed as the ratio of the fracture energy at a given time $W_r(t)$ to that ($W_{12}$) after 12 hours) obtained from the rolling torque measurements[48].

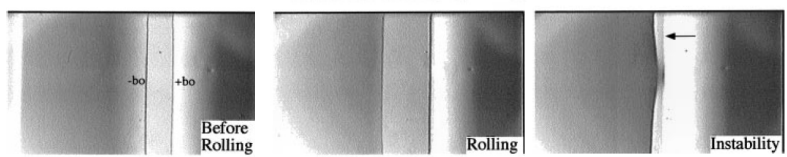

**Figure 13.** Elastic instability[48] observed with the rolling of cylinder on a surface that shows age dependent adhesion. On the figure marked "Instability", the trailing edge, which is at the junction of high and low adhesion regions, undergoes an instability and the hemicylinder exhibits fast rolling before the trailing edge passes to the region of lower adhesion.

These experiments provide evidence that not only sliding friction, but also rolling friction is affected by interfacial processes and time of contact.

## Summary and Perspective

We have tried to present a brief review of the old and new concepts of rubber friction that show certain parallelism with the sliding friction of gels. These results should also be relevant to various thin film studies involving grafted polymers. Because of somewhat limited focus of the review, many important studies could not be reviewed here, such as the studies involving the friction of polymer brush in solvents[49] or the chain pull-out effects[50-52] that play important roles in the friction of grafted brush against rubbery networks. However, the brief review of the friction of crosslinked networks against impenetrable surfaces raises several important questions that have not yet found clear answers. The various models used to describe rubber friction differ considerably from each other in their essential physical foundation. For example, the model of Schallamach considers the cyclic processes of attachment and detachment of the polymer chains at the interface, whereas Persson's model is based on thermally activated fluctuation of the stress segments that never de-adheres from the surface. Another model[22] describes friction as a stress-biased diffusion of polymer chains in contact with the surface. Definitive experiments are needed to examine the interface and friction measurements simultaneously, to address the detailed questions of dynamics of polymer chains as well as its interactions with surfaces. Some important developments in this direction are taking place, especially with the advent of sum-frequency generation spectroscopy[53]. However, more developments to interrogate the interface are needed. An important question that needs to be resolved is the aging behavior of polymer chains at the interface and its relationship to the adhesive and tribological interactions that prevail at the interface.